\documentclass[12pt]{article}
\usepackage{epsfig}
\usepackage{graphicx}
\usepackage{a4}
\usepackage{amsmath}
\usepackage{latexsym}
\usepackage{cite}
\usepackage{lineno}
\usepackage{color}
\usepackage{colordvi}

\graphicspath{{Pics/}}
\DeclareGraphicsExtensions{.eps,.ps}

\usepackage{pslatex}
\usepackage[latin1]{inputenc}
\usepackage[T1]{fontenc}

\usepackage{amssymb}

\usepackage{url}

\newcommand{\msbar}{$\overline{\text{MS}}\, $}

\begin{document}

\begin{titlepage}
\noindent
%Draft 2.0  \hfill 04 March 2015\\
\\
DESY 15-034 \hfill  2015\\
\\

\vspace{1.3cm}

\begin{center}
  {\bf 

\large

Impact of heavy-flavour production cross sections measured by the LHCb 
experiment on parton distribution functions at low x}
  \vspace{1.5cm}

  {\large
    PROSA Collaboration
  }\\

  \vspace{1.2cm}

\end{center}

\noindent
O.~Zenaiev$^{1}$, A.~Geiser$^{1}$, K.~Lipka$^{1}$, 
J.~Bl\"umlein$^{1}$, A.~Cooper-Sarkar$^{2}$, \mbox{M.-V.~Garzelli}$^{3}$, M.~Guzzi$^{4}$, O.~Kuprash$^{1}$,
\mbox{S.-O.~Moch}$^{3}$, P.~Nadolsky$^{5}$, R.~Placakyte$^{1}$, K.~Rabbertz$^{6}$, I.~Schienbein$^{7}$, P.~Starovoitov$^{1}$\\

\noindent

{\footnotesize{
\noindent
$^{1}$DESY Hamburg \& Zeuthen, Germany,
$^{2}$University of Oxford, UK,
$^{3}$Universit\"at Hamburg, Germany,
$^{4}$School of Physics and Astronomy, the University of Manchester, UK,
$^{5}$Southern Methodist University, Dallas, Texas, USA,
$^{6}$Karlsruher Institut f\"ur Technologie, Germany,
$^{7}$LPSC Grenoble, France.\\

}

}

  \vspace{2.4cm}
\begin{center}
\large
{\bf Abstract}
\vspace{-0.2cm}
\end{center}
The impact of recent measurements of heavy-flavour production in deep inelastic $ep$ scattering and in $pp$ collisions
on parton distribution functions is studied in a QCD analysis in the 
fixed-flavour number scheme at next-to-leading order. 
Differential cross sections of charm- and beauty-hadron production measured by
LHCb are used together with inclusive and heavy-flavour production cross sections in deep inelastic 
scattering at HERA. The heavy-flavour data of the LHCb experiment impose additional constraints 
on the gluon and the sea-quark distributions at low partonic fractions $x$ of the proton momentum, 
down to $x \sim 5 \times 10^{-6}$. This kinematic range is currently not 
covered by other experimental data in perturbative QCD fits.
\vfill
\end{titlepage}

%
% ----------------------------------------------------------------------------------------------
%
\newpage
\section{Introduction}
\label{sect:int}

Understanding the nucleon structure is one of the fundamental tasks of modern particle 
physics. In quantum chromodynamics (QCD), the structure of the nucleon is described by parton 
distribution functions (PDFs), which, in collinear factorisation, represent probability densities to find 
a parton of longitudinal fraction $x$ of the nucleon momentum at a factorisation scale $\mu_f$. 
%Renormalisation group equations, e.g. DGLAP evolution equations~\cite{scaling_violations}, \cite{dglap},
%predict the dependence of the PDFs on $\mu$ in perturbative QCD (pQCD). 
The scale evolution of the PDFs is uniquely predicted by the renormalisation group equations for factorisation~\cite{scaling_violations, dglap}.
The $x$-dependence
cannot be derived from first principles and must be constrained by experimental measurements. 
The precision of the PDFs is of key importance for interpreting the measurements in hadronic collisions. 
In particular, the uncertainty of the proton PDFs must be significantly reduced in order to improve the
accuracy of theory predictions for Standard Model (SM) processes at the LHC. 

Deep inelastic lepton-proton scattering (DIS) experiments cover a broad range 
in $x$ and $\mu_f$. In the perturbative regime,  
a wide $x$-range of $10^{-4}<x\lesssim 10^{-1}$ is probed by the data of the H1 and ZEUS experiments 
at the HERA collider \cite{heradis}. These measurements impose the tightest constraints on the existing PDFs. 
However, additional measurements are necessary for a better flavour separation and to constrain the kinematic ranges 
of very small and very high $x$, where the gluon distribution is poorly known. A better constraint on the high-$x$ gluon 
is needed for an accurate description of the SM backgrounds in searches for new particle production at high masses or 
momenta. Significant reduction of the uncertainty of the low-$x$ gluon distribution is important for studies of 
parton dynamics, non-linear and saturation effects. Furthermore, precision of the gluon distribution at low $x$ has 
implications in physics of atmospheric showers, being crucial for cross-section predictions of high-energy neutrino 
%production~\cite{cooper_sarkar_2011}and for calculations of prompt fluxes of lepton production in 
%the atmosphere~\cite{pasquali}.
DIS interaction~\cite{cooper_sarkar_2011} and for calculations of prompt lepton fluxes in the atmosphere~\cite{pasquali}.
\begin{figure}[h]
\center
\epsfig{file=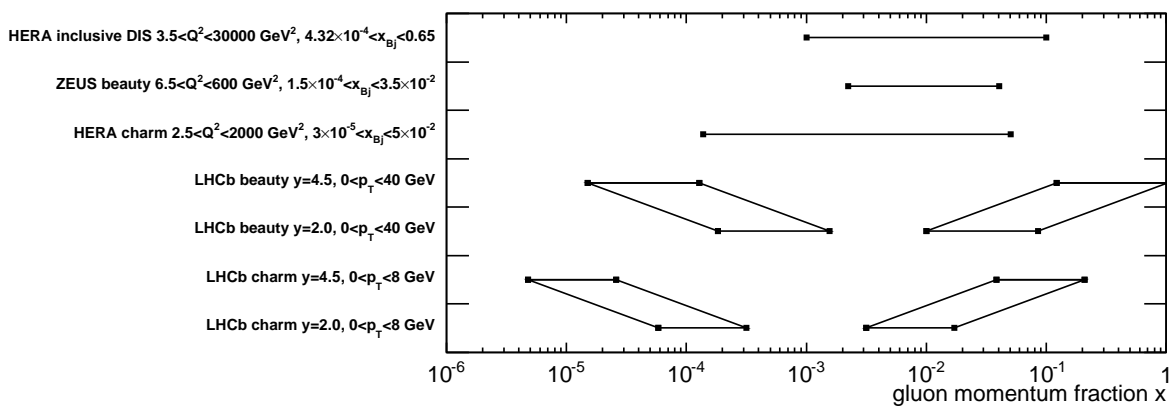,width=1.0\textwidth}
\caption{Kinematic range in $x$ for the gluon density 
covered by measurements at HERA and LHCb. 
For the HERA inclusive DIS data, the $x$ range is indicated, where the gluon PDF 
uncertainties are less than 10\% at $\mu_f^2 = 10$ GeV$^2$. For the LHCb data, the 
upper (lower) edge of the box refers to the indicated upper (lower) end of the rapidity, 
$y$, range of the heavy-hadron production.}
\label{kinematics} 
\end{figure}

Heavy-flavour measurements of the LHCb Collaboration~\cite{lhcb_charm,lhcb_beauty} at the LHC 
probe the very forward range of the heavy-hadron rapidity $y$ and are sensitive to the gluon 
PDF at low $x$, as schematically shown in Fig.~\ref{kinematics}. For this illustration, 
in the calculation of the kinematics of heavy-quark production at HERA, the 
leading order (LO) relation is used for the typical gluon $x$ in boson-gluon fusion,
$x=x_{Bj}(1+ \frac{4 m_Q^2}{Q^2})$, where $x_{Bj}$ denotes the Bjorken 
scaling variable, $m_Q$ is the heavy-quark mass, and $Q^2$ is the 
virtuality of the exchanged electroweak boson. 
In the case of heavy-quark production at LHCb, the LO formula 
$x=e^{\pm y}\frac{\sqrt{p_T^2+m_Q^2}}{E_p}$, assuming $p_z=0$ in the parton-parton rest frame, is 
applied. Here, $p_T$ and $p_z$ represent the transverse and longitudinal momenta of the heavy quark, 
respectively, and $E_p$ is the proton beam energy.

Heavy-flavour production in proton-proton collisions at LHC is dominated by the 
gluon-gluon fusion process. Therefore the LHCb measurements of charm~\cite{lhcb_charm} and beauty~\cite{lhcb_beauty} 
production in the forward region $2.0 < y < 4.5$ probe the gluon distribution at 
$5\times 10^{-6} \lesssim x \lesssim 10^{-4}$, a region which is not accessible with HERA data. 
Note that the LHCb data are sensitive to the product of gluon densities in two non-overlapping low and 
medium-to-high $x$ ranges, as illustrated in Fig.~\ref{kinematics}. Since the medium range is already well 
constrained by HERA data, which furthermore bridge the gap between 
the two LHCb ranges,
the major impact of the LHCb heavy-flavour measurements is expected at $5\times 10^{-6} \lesssim x \lesssim 10^{-4}$.
The advantage of using heavy-flavour data 
is that the charm and beauty masses provide hard 
scales for the perturbative QCD expansion all the way down to their 
production threshold.
To estimate the impact of the LHCb measurement of charm and beauty production on the gluon distribution at low $x$, 
these data are included in a QCD analysis together with the inclusive DIS~\cite{heradis} and heavy-flavour
production~\cite{herac,heracb} cross sections measured at HERA. 
%
%-----------------------------------------------------------------------------------------------------
%
\section{Experimental data used in the QCD analysis}
\label{sect:exp}

The main objective of the present QCD analysis is to demonstrate the constraining power of the measurements of heavy-flavour production in DIS and $pp$ collisions for the determination of 
the PDFs of the proton. The measurements of charm and beauty production at HERA and LHCb, together with 
the combined HERA inclusive cross-section measurements, are used in a 
next-to-leading order (NLO) perturbative QCD (pQCD) analysis.

Neutral current (NC) and charged current (CC) inclusive DIS cross sections in $ep$ scattering are directly
sensitive to the valence and sea-quark distributions and probe the 
gluon distribution through scaling 
violations \cite{scaling_violations}. 
HERA measurements of the NC and CC cross sections in DIS at 
a centre-of-mass energy $\sqrt{s}$~=~320~GeV have been combined taking into account systematic correlations~\cite{heradis}. 
This combined data set contains the complete information on inclusive DIS cross sections published by the H1 and 
ZEUS Collaborations based on data collected in the years 1994-2000, and has been used for the determination 
of the PDF set HERAPDF1.0~\cite{heradis}. 
The kinematic range of the NC data is $6 \times 10^{-7} \leq x_{Bj} \leq 0.65$, $0.045 \leq Q^2 \leq 30000$ GeV$^2$. 
% where $Q^2$ is the virtuality of the exchanged electroweak boson. 
The CC cross sections span the kinematic range of $1.3 \times 10^{-2} \leq x_{Bj} \leq 0.40$ and $300 \leq Q^2 \leq 30000$ 
GeV$^2$. These combined NC $e^\pm p$ and CC $e^\pm p$ cross sections represent 
the basis for all PDF determinations.

In $ep$ scattering, charm and beauty quarks are produced predominantly in the photon-gluon fusion process 
which provides a direct probe of the gluon distribution in the proton. 
Measurements of open-charm production 
cross sections in DIS at HERA from the H1 and ZEUS Collaborations have been 
combined \cite{herac}. Cross sections  for charm 
production were obtained in the kinematic range of $2.5 \leq Q^2 \leq 2000$ GeV$^2$ and $ 3 \times 10^{-5} \leq x_{Bj} \leq 5 \times 10^{-2}$. The combination method accounts for the correlations of the systematic uncertainties among the
different data sets. These combined measurements were used to improve 
constraints on the gluon distribution and 
to determine the charm-quark mass~\cite{herac}. The charm reduced cross sections determined 
as a function of $Q^2$ and $x_{Bj}$ are used in the present analysis 
together with all provided details on the systematic 
correlations. In addition, cross sections for the production of $b$ quarks in $ep$ scattering, as 
measured by the ZEUS Collaboration~\cite{heracb} are used in the present analysis. 
These data correspond to an integrated luminosity of 354 pb$^{-1}$ and cover the kinematic range 
of $5 < Q^2 <1000$ GeV$^2$. The $b$- and $c$-quark content in the events with at least one 
jet have been extracted using the invariant mass of charged tracks associated with secondary vertices and 
%the decay-length significance of these vertices and the $b$-quark mass has been measured \cite{heracb}. 
lifetime information, and the $b$-quark mass has been measured \cite{heracb}. 
In the present analysis, the $b$-quark production data are used mainly to improve constraints on the $b$-quark mass.

For additional constraints on the gluon distribution at low $x$ the differential cross sections 
of charm and beauty production in $pp$ collisions at $\sqrt{s}=7$ TeV from the LHCb experiment are 
used for the first time. The measurement of charm production~\cite{lhcb_charm} is based on data corresponding 
to an integrated luminosity of 15 nb$^{-1}$. Charm production is identified through the full reconstruction 
of decays of the charmed hadrons\footnote{Charge conjugation is always implied for charm and beauty hadrons.} 
$D^0$, $D^{+}$, $D^{*+}$, $D_s^+$ and $\Lambda_c^{+}$. 
The cross sections are measured 
as a function of the transverse momentum, $p_T$, and rapidity, $y$, of the reconstructed hadrons. 
The LHCb data on $B$-meson production in $pp$ collisions~\cite{lhcb_beauty} correspond to an integrated 
luminosity of 0.36 fb$^{-1}$. The $B^+$, $B^0$ and $B_s^0$ mesons are reconstructed in exclusive decays 
mainly involving $J/\psi$ final states. Correlations between the experimental systematic uncertainties 
are accounted for as described in the original publications. An uncorrelated systematic uncertainty is 
obtained for each distribution by subtracting the correlated uncertainties from the total ones. 
The 3.5\% luminosity uncertainty is treated as correlated between the measurements of charm and beauty production.
In the present analysis,
the normalised cross sections, $\frac{{\rm d}\sigma}{{\rm d}y} / \frac{{\rm d}\sigma}{{\rm d}y_0}$, 
for charm and beauty production are calculated from the absolute measurements published by LHCb and are used in the 
QCD analysis, with $\frac{{\rm d}\sigma}{{\rm d}y_0}$ being the cross section in the center bin, $3 < y < 3.5$, of 
the measured rapidity range in each $p_T$ bin. The uncorrelated experimental uncertainty 
on $\frac{{\rm d}\sigma}{{\rm d}y_0}$ is propagated as a correlated uncertainty to the respective complementary 
rapidity bins. The QCD analysis is performed by using both, absolute or normalised, representations of the LHCb measurements, alternatively.
\section{Theoretical predictions for heavy-flavour production}
\label{sec:th}

In the QCD analysis, the experimental measurements are confronted with 
corresponding theoretical predictions. The theoretical predictions for 
charm and beauty production in both $ep$ and $pp$ collisions are obtained at 
NLO in the fixed-flavour number scheme (FFNS). 
This scheme and its applicability to HERA measurements is discussed in 
detail in Ref.~\cite{herac} and references therein. Predictions for HERA data 
are obtained by following the approach of the ABM group at NLO using its 
implementation in OPENQCDRAD \cite{openqcdrad} in the 
framework of HERAFitter~\cite{herafitter}. The number 
of active flavours is set to $N_f = 3$, and the renormalisation and factorisation scales (pQCD scales) for heavy flavour 
production are chosen as $\mu_r = \mu_f=\sqrt{Q^2+4m_Q^2}$, where $m_Q$ denotes 
the pole mass of $c$ or $b$ quarks\footnote{The pole mass is used for consistency with the $pp$ predictions,
since $\overline{\rm MS}$ running mass predictions are not available for LHCb.}.
For the light-flavour contributions to the inclusive DIS cross sections,
the pQCD scales are set to $\mu_r = \mu_f = Q$. 

Theoretical predictions for heavy-quark production in $pp$ collisions are 
obtained using the massive NLO calculations~\cite{MNR,Beenakker:1988bq,MNRSingleCalc} in the FFNS, 
also available as part of the Mangano-Nason-Ridolfi (MNR) calculations~\cite{MNRFullyExclusive}. The pQCD scales are chosen as 
$\mu_{r,f} = A_{r,f}^{c,b}\mu_0$, with $\mu_0=\sqrt{p^2_T+m^2_Q}$ and $A_{r,f}^{c,b}$ being coefficients 
for $c$ and $b$ quarks, which are discussed in the following. 
These predictions were used successfully for beauty production in $p \bar p$ 
collisions at the $Sp\bar pS$ \cite{UA1} and the Tevatron\footnote{Provided
that fragmentation and other uncertainties are properly 
treated \cite{pertfrag}. Note that the NLO+NLL (FONLL) \cite{FONLL} 
calculations used there, and also used by LHCb \cite{lhcb_charm,lhcb_beauty}, 
slightly reduce 
the cross sections at high transverse momenta with respect to the pure NLO 
calculation, 
while they are identical 
at low transverse momenta \cite{FONLL}. The Tevatron data are well described 
by FONLL~\cite{pertfrag}.
This conclusion is also applicable to the NLO calculations 
\cite{MNRSingleCalc} used here, since these were used as input for the 
NLO part of the FONLL calculations.
The claim in \cite{bTevatron} that the NLO predictions undershoot 
the data while FONLL describes them must thus be attributed to 
parametrisations beyond the perturbative 
part of the calculations, as illustrated, e.g., in Fig. 7 of~\cite{Kniehl:2008zza}.}~\cite{bTevatron}. 
They are conceptually very similar to the Frixione-Mangano-Nason-Ridolfi (FMNR)
predictions~\cite{FMNR} employed for heavy-flavour photoproduction at HERA \cite{HERAphot}. 

The cross-section predictions for heavy-flavoured hadron production not 
only depend on the kinematics of the heavy-flavour production mechanism, but 
also on the fragmentation of the heavy quark into a particular final-state hadron. 
There is no final-state factorisation scale in the FFNS since
collinear logarithms of the heavy-quark mass are included in fixed-order 
perturbation theory.
The calculations in 
\cite{Beenakker:1988bq,MNR,MNRSingleCalc,Beenakker:1990ma} describe 
the production of an on-shell heavy quark. Near the kinematic threshold, 
the transition of the heavy quark 
into the observed heavy-flavoured hadron can be taken into account by 
multiplying the cross section with the appropriate branching fraction. 
This leads to an excellent description of $B$- and $D$-meson production measurements at the Tevatron 
and the LHC from $p_T=0$ up to $p_T \sim 4 m$~\cite{Kniehl:2015fla,Kniehl:2012ti}.
The scope of these calculations can be extended by convoluting the 
heavy-quark production cross section with a suitable scale-independent 
fragmentation function describing the hadronisation of the heavy quark. 
The implementation of the convolution is not unique once the quark and 
hadron masses are taken into account, and leads to a potentially $p_T$-dependent
modelling uncertainty which 
is, however, small compared to the scale-choice uncertainty at NLO. 
This fragmentation function is used on a purely phenomenological basis, since 
it does not strictly appear in the context of a factorisation theorem, and therefore it has to be extracted from data.
It depends on the order of the perturbation series but is generally assumed 
to be otherwise universal. Its main effect is to lower the theoretical 
predictions at large $p_T$. Typical parametrisations used in the literature 
are those by Peterson et al.~\cite{Peterson:1983ak} depending on one 
parameter $\varepsilon$ and by Kartvelishvili et al. \cite{kartv} 
depending on one parameter $\alpha_K$. 

For the HERA measurements, the fragmentation functions and their uncertainties are considered and accounted for 
in the original publications~\cite{herac, heracb}. The measurements of LHCb are provided as hadron-production 
cross sections and the fragmentation functions have to be applied explicitly in order  
to use these data in the QCD analysis. In addition, fragmentation fractions
describing the probability of a quark to fragment into a particular hadron have to be applied.  
The fragmentation fractions for $c$-flavoured hadrons are taken from~\cite{44} and for $b$-flavoured hadrons from~\cite{98}.

So far, no fragmentation measurements were performed in $pp$ collisions. 
Because of similarities of the $c$-quark production kinematics at HERA and 
LHCb, the Kartvelishvili fragmentation function~\cite{kartv}
with $\alpha_K=4.4\pm1.7$, 
as obtained from corresponding HERA measurements~\cite{h1frag,zeusfrag} 
extracted for the NLO FFNS scheme, 
is applied for predictions of the LHCb 
measurements of charm-hadron production.
The fragmentation is performed in the laboratory frame 
by rescaling  the quark three-momentum with the energy of the produced hadron being calculated 
using the hadron mass. 
This procedure is used for $D^+$ and $D^+_s$ mesons, and for $\Lambda_c^+$ baryons. 
For $D^0$- and $D^+$-meson production, the contribution from $D^{*+}$ and $D^{*0}$ mesons is treated as described in~\cite{106}. 
For beauty production, the value $\alpha_K= 11\pm 4$ is used for all $b$-flavored hadrons, corresponding 
to measurements at LEP~\cite{103}. 

The fragmentation-fraction uncertainties are assigned to the 
measurements and are treated as correlated, while the uncertainties arising 
from the variations of assumptions on the fragmentation functions are treated in the form of variations of the 
theory predictions in the QCD fit. 

\section{Details of the QCD analysis}

The open source QCD fit framework for PDF determination HERAFitter~\cite{herafitter}, version 1.0.0, is used. 
The partons are evolved by using the QCDNUM program~\cite{qcdnum}. The analysis-specific modifications to HERAFitter 
address the heavy-flavour treatment as follows. The massive fixed-flavour number scheme~\cite{ffns} with the number 
of flavours $N_f=3$ is used for the treatment of heavy-flavour contributions. The calculation of one-particle inclusive 
heavy-quark production cross sections in hadron collisions at NLO according to~\cite{MNRSingleCalc} is implemented 
by using original routines from the MNR code~\cite{MNRCode}. The results agree with those obtained with the original 
MNR code at a level of accuracy below 1\%. 

The 3-flavour strong coupling constant in the NLO \msbar scheme is set to 
$\alpha_S(m_Z)^{N_f=3} =0.1059\pm 0.0005$, which corresponds to
the world average value of $\alpha_S(m_Z)^{N_f=5}= 0.1185\pm0.0006$,
using two-loop evolution equations~\cite{qcdnum}.

The $Q^2$ range of the inclusive HERA data is restricted to $Q^2>Q^2_{min}$ = 3.5 GeV$^2$. 
The procedure for the determination of the PDFs follows the approach used in the HERAPDF1.0 QCD fit~\cite{heradis}. 
The following independent combinations of parton distributions are chosen in the fit procedure at the initial scale 
of the QCD evolution $Q^2_0= 1.4$ GeV$^2$: the valence-quark distributions $xu_{\textrm{v}}(x)$, 
$xd_{\textrm{v}}(x)$, the gluon distribution $xg(x)$ and the $u$-type and $d$-type 
anti-quark distributions (which are identical to the sea-quark distributions), 
$x\overline{\textrm{U}}(x)$, $x\overline{\textrm{D}}(x)$, where
$x\overline{\textrm{U}}(x) = x\overline{u}(x)$ and $x\overline{\textrm{D}}(x) = x\overline{d}(x) + x\overline{s}(x)$. 
At the scale $Q_0$, the parton distributions are represented by
\begin{eqnarray}
x u_\textrm{v}(x) &=& A_{u_{\textrm{v}}} ~  x^{B_{u_{\textrm{v}}}} ~ (1-x)^{C_{u_{\textrm{v}}}} ~(1+E_{u_{\textrm{v}}} x^2) ,
\label{eq:uv}\\
x d_\textrm{v}(x) &=& A_{d_{\textrm{v}}} ~ x^{B_{d_{\textrm{v}}}} ~ (1-x)^{C_{d_{\textrm{v}}}},  
\label{eq:dv}\\
x \overline {\textrm{U}}(x) &=& A_{\overline {\textrm{U}}} ~ x^{B_{\overline {\textrm{U}}}} ~ (1-x)^{C_{\overline {\textrm{U}}}},
\label{eq:Ubar}\\
x \overline {\textrm{D}}(x) &=& A_{\overline{\textrm{D}}} ~ x^{B_{\overline{\textrm{D}}}} ~ (1-x)^{C_{\overline{\textrm{D}}}}, 
\label{eq:Dbar}\\
x g(x) &=& A_{g} ~ x^{B_{g}} ~ (1-x)^{C_{g}} 
- A'_{g} ~ x^{B'_{g}} ~ (1-x)^{C'_{g}}.  
\label{eq:g}
\end{eqnarray}
The normalisation parameters $A_{u_{\textrm{v}}}$, $A_{d_\textrm{v}}$, $A_g$ are determined by the QCD sum 
rules, the $B$ parameters are responsible for the small-$x$ behaviour of the PDFs, and the parameters $C$ 
describe the shape of the distribution as $x \to 1$. A flexible form for the gluon distribution is 
adopted with the choice of $C'_g=25$ motivated by the approach of the MSTW group~\cite{Thorne:2006qt,Martin:2009ad}.
The $s$-quark distribution is defined through
$x$-independent strangeness fraction, $f_s$, of the $d$-type sea, $x\overline{s} = f_sx\overline{D}$ at $Q^2_0$, 
%$f_s = \bar{s}/( \bar{d} + \bar{s})$ 
%where $f_s=0.31\pm0.08$ as in the analysis of Ref.~\cite{Martin:2009ad}. 
where $f_s=0.31^{+0.19}_{-0.08}$ as in the analysis of~\cite{Martin:2009ad}, 
including the recent complementary measurement~\cite{atlasfs}. 
Additional constraints $B_{\overline{\textrm{U}}} = B_{\overline{\textrm{D}}}$ and 
$A_{\overline{\textrm{U}}} = A_{\overline{\textrm{D}}}(1 - f_s)$ are imposed, with $x\bar{u} \to x\bar{d}$ as $x \to 0$.
The analysis is performed by fitting the remaining 13 free parameters in Eqs.~(\ref{eq:uv}--\ref{eq:g}). 

The PDF parameters are determined in HERAFitter by minimisation of a $\chi^2$-function 
taking into account correlated and uncorrelated uncertainties~\cite{HERAFitterPaper} of the measurements. 
Systematic uncertainties are assumed to be proportional to the central prediction values, 
whereas statistical uncertainties scale with the square root of the predictions. 
Correlated uncertainties are treated using nuisance parameter representation~\cite{HERAFitterPaper}. 
To minimise biases arising from the likelihood transition to $\chi^2$ when the scaling 
of the errors is applied, a logarithmic correction is added to the $\chi^2$-function~\cite{lntermH1}.

The heavy-quark masses are left free in the fit. They are well constrained by 
the measurements of charm and beauty production in DIS and the fitted values (see Table~\ref{tab:fitpar} in the Appendix~\ref{sec:Appendix}) are consistent with the ones obtained in the corresponding HERA 
analyses~\cite{herac,heracb} within the intrinsic theoretical systematic 
uncertainty of the pole mass definition~\cite{poleunc}.  

The QCD analysis is performed twice using either absolute or normalised differential cross sections of heavy-flavour 
production from LHCb measurements, as defined in Section \ref{sect:exp}. 
The implementation of the theory calculations~\cite{thesis} as 
described in Section~\ref{sec:th} allows the pQCD scales, i.e. the parameters $A_{r,f}^{c,b}$, and the values for the 
pole mass of the heavy quarks to be changed at each fit iteration. 

In the QCD analysis using the normalised LHCb measurements, the pQCD scales are fixed to $A_r=A_f=1$ for 
the central result. The scale dependence is studied by varying the pQCD scales independently such that 
$0.5\leq A_r,A_f \leq 2$. 
$A^c$ and $A^b$ are always varied simultaneously. The resulting 
scale dependence is small, since it is largely absorbed by the normalisation, as illustrated in Appendix~\ref{sec:Appendix}.

In the variant of the fit using the absolute LHCb cross sections, the scale dependence of the predicted 
cross section is the dominant theoretical uncertainty. The same scale choice and 
variation procedure, as applied for the variant of the fit using the normalised LHCb measurements, 
leads to unacceptably high $\chi^2$ values of the respective fits~\cite{thesis}. 
Therefore, the four scales technically are treated as independent fully 
correlated systematic uncertainties for the central result. 
Since the pQCD scales are not physical parameters, the related uncertainties are 
not obtained from the fit. 
Instead, the effect of the scale choice on the other fitted parameters is
evaluated by an independent variation of $A_f$ in the range $0.5< A_f^c=A_f^b < 2$ with $A_r^c$ and $A_r^b$ as 
free parameters, or $A_r$ in the range $0.25 < A_r^c = A_r^b < 1$ with $A_f^c$ and $A_f^b$ being free parameters. 
For the variation $A^c_f=A^b_f=0.5$, a cut $p_T>2$ GeV is applied for the 
charm LHCb data to ensure that the factorisation scale is above $1$ GeV${}^2$, since this is technically required in the QCDNUM. 
This procedure ensures an acceptable fit quality for all variations~\cite{thesis}, as required for a meaningful extraction 
of the other uncertainties. Because of the unconventional scale treatment the fit using absolute cross sections is considered 
to be a cross check.
%
%-----------------------------------------------------------------------------------------------------
\section{PDF uncertainties}
\label{sec:uncertainties}
The PDF uncertainties are estimated following the approach of HERAPDF1.0~\cite{heradis} in 
which experimental, model, and parametrisation uncertainties are taken into account. 
Experimental uncertainties are evaluated using the Hessian method~\cite{HERAFitterPaper}. 
A tolerance criterion of $\Delta \chi^2 = 1$ is adopted for defining the fit uncertainties that originate from the experimental 
uncertainties of the measurements included in the analysis. 

Model uncertainties arise from the variations in the values assumed for $Q^2_\mathrm{min}$ imposed on the HERA data, 
which is varied in the interval $2.5\le Q^2_\mathrm{min} < 5.0$ GeV$^2$; the fraction of strange quarks,  
varied in the range $0.23<f_s<0.50$ and the value of the strong coupling, varied in the 
range $0.1054<\alpha_S(m_Z)^{N_F=3}<0.1064$. The pQCD scales for heavy-quark production in DIS are varied 
simultaneously by a factor of 2 up and down for both, charm and beauty. 
For the fits with the LHCb data, the model uncertainties include theoretical uncertainties for 
the cross section predictions for heavy-flavoured hadron production, arising from variation of the pQCD scales and of 
the fragmentation parameters, as described in Section~\ref{sec:th}. Uncertainties, arising from these model variations are 
referred to as MNR uncertainties in the following.

The parametrisation uncertainty is estimated similarly to the HERAPDF1.0 procedure: for all parton densities, 
additional parameters are added one by one in the functional form of the parametrisations in Eqs.~(\ref{eq:uv}--\ref{eq:g}), 
in a similar way as described in~\cite{heradis,herac,heracb}. 
Furthermore, the starting scale is varied to $Q^2_0=1.9$ GeV$^2$. 
The parametrisation uncertainty is constructed as an envelope built from the maximal differences between the 
PDFs resulting from all the parametrisation variations and the central fit at each $x$ value. The total 
PDF uncertainty is obtained by adding experimental, model and parametrisation uncertainties in quadrature. 
%----------------------------------------------------------------------------------------------------
%
\section{Results}
\label{sec:results}
\begin{figure}[ht]
\center
\epsfig{file=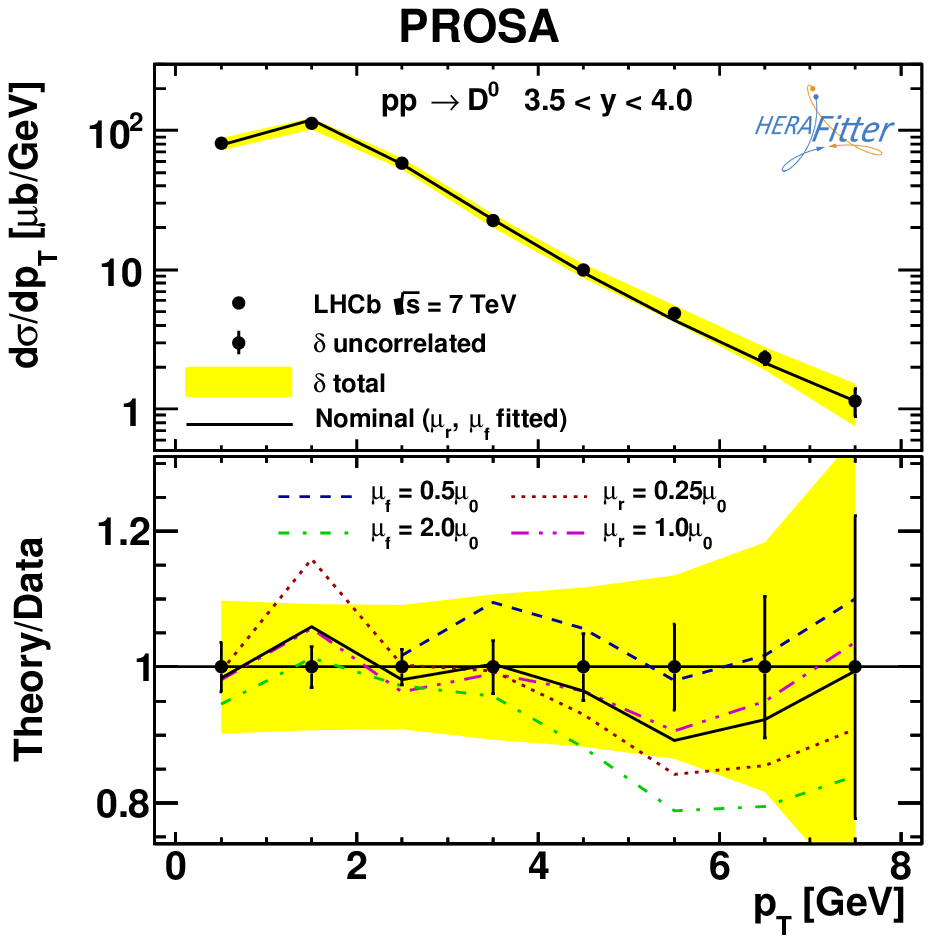,width=0.495\textwidth}
\epsfig{file=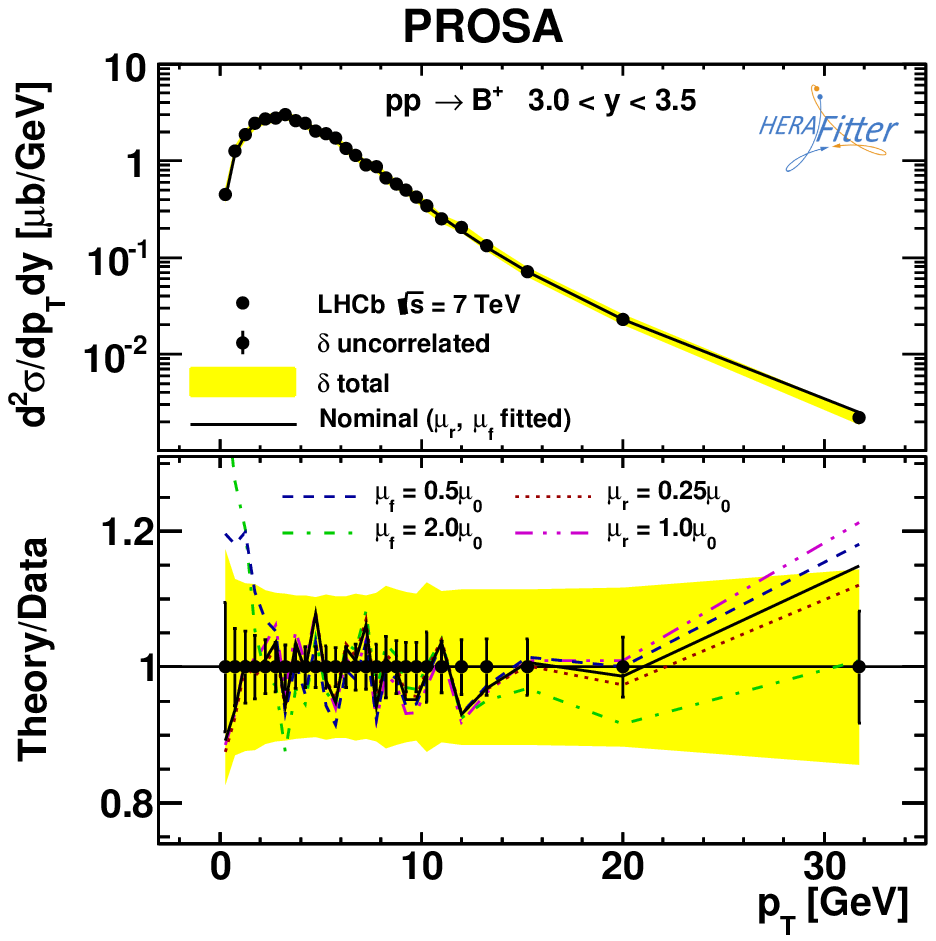,width=0.495\textwidth}
\caption{Data to theory comparison for a representative subset of the LHCb absolute cross sections for production of $D^0$ 
mesons for $3.5<y<4.0$ (left) and of $B^{+}$ mesons for $3.0<y<3.5$ (right). In the bottom panels the ratios theory/data 
for the nominal variant of the fit and the scale variations are shown. For demonstration purpose, correlated shifts 
for data points obtained in the fit using nuisance parameters are applied to theoretical predictions. Uncorrelated 
uncertainties for data points are shown as they are rescaled in the fit, while total uncertainties are shown as not rescaled.}
\label{fig:data_theory_abs} 
\end{figure}

\begin{figure}[ht]
\center
\epsfig{file=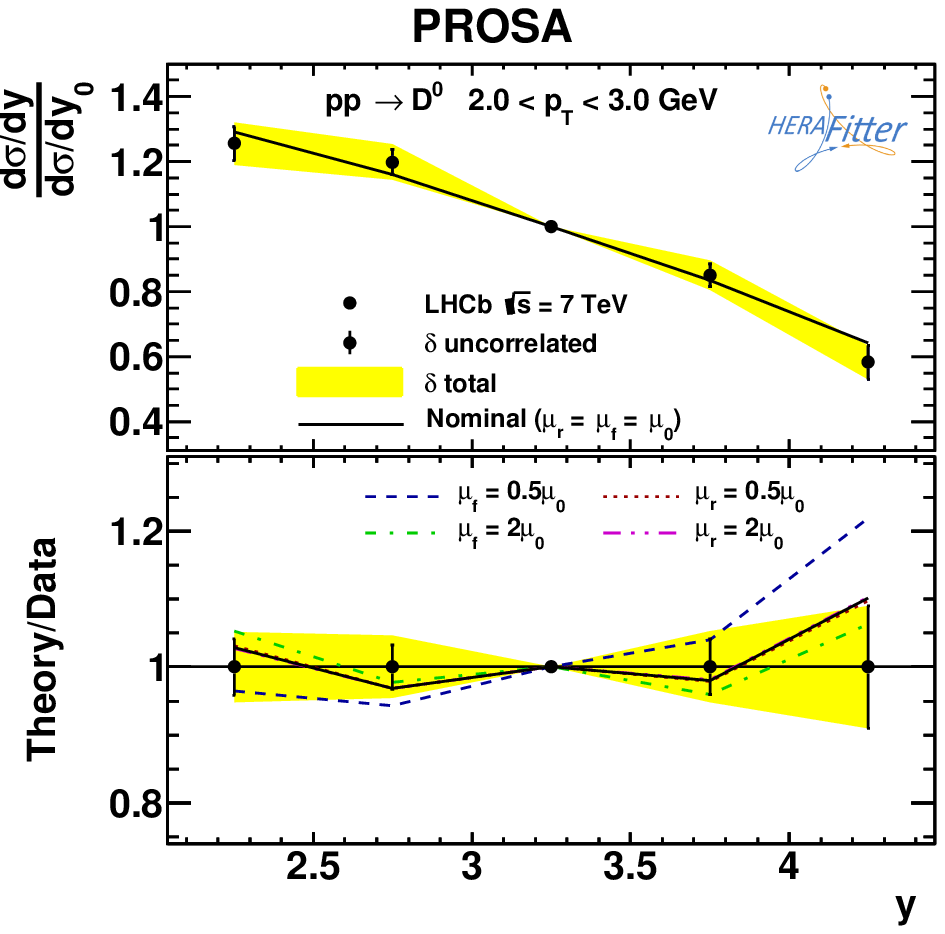,width=0.495\textwidth}
\epsfig{file=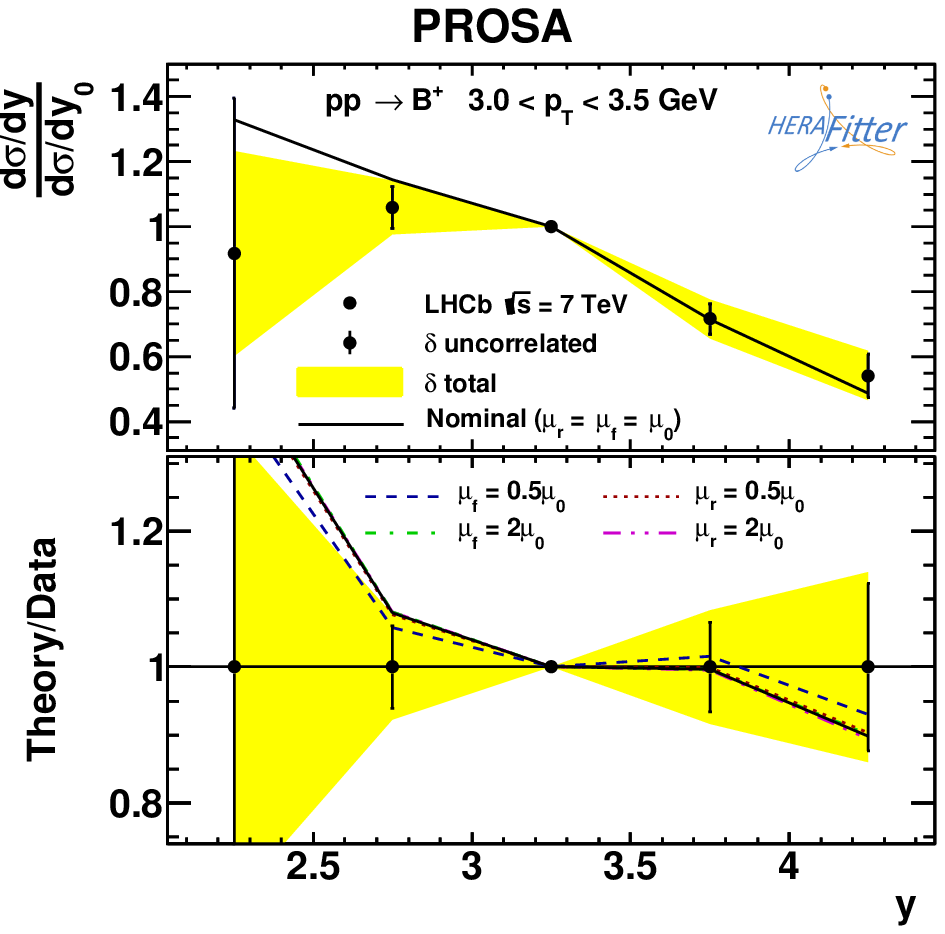,width=0.495\textwidth}
\caption{Data to theory comparison for a representative subset of the LHCb normalised cross sections for production of $D^0$ mesons for $2.0 < p_T < 3.0$~GeV (left) and of $B^{+}$ mesons for $3.0 < p_T < 3.5$~GeV (right). 
The central rapidity bins are fixed to $1$ by the definition of the normalised cross sections. 
In the bottom panels the ratios theory/data for the nominal variant of the fit and the scale 
variations are shown. For demonstration purpose, correlated shifts for data points obtained in the fit using nuisance 
parameters are applied to theoretical predictions. Uncorrelated uncertainties for data points are shown as they are 
rescaled in the fit, while total uncertainties are shown as not rescaled.}
\label{fig:data_theory_norm} 
\end{figure}

In Fig.~\ref{fig:data_theory_abs}, 
the absolute cross sections for $D^0$- and $B^{+}$-meson production in $pp$ collisions are shown for one 
representative rapidity bin and are compared to the theory predictions as used in the QCD analysis. A significant scale 
dependence is observed. The normalised cross sections for a representative $p_T$ bin of the same data set are compared 
to the respective theory predictions in Fig.~\ref{fig:data_theory_norm}. The advantage of using the normalised cross 
section is a significant reduction of the scale dependence of the theoretical prediction, retaining the sensitivity of 
the cross sections to the gluon distribution. The reduction of the uncertainty due to scale variation is 
related to the fact that the scale choice affects mostly the normalisation but only to some extent the 
shape of heavy-quark production kinematics, as demonstrated in Fig.~\ref{fig:nlovar_c}, \ref{fig:nlovar_b} 
in the Appendix~\ref{sec:Appendix}.


The fit quality, represented by the total and partial values of $\chi^2$ divided by the number 
of degrees of freedom, $n_{\rm dof}$, for both variants of the QCD analysis is presented in Table~\ref{tab:chi}.
When the normalised LHCb measurements are used in the QCD analysis, 
$n_{\rm dof}$ is appropriately reduced for the respective data sets. 
The fitted parameters are presented in Table~\ref{tab:fitpar} in the Appendix~\ref{sec:Appendix}.

\begin{table}[t]
\begin{center}
\begin{tabular}{|l|c|c|} \hline
\small  Representation of the LHCb measurements  & absolute  &  normalised\\ \hline 
Global $\chi^2 / n_{\rm dof}$            & 1073 / 1087    &   958  /  994\\
Global $\chi^2$ $p$-value               & 0.61           &   0.79    \\ 
$\chi^2$-contribution from correlated uncertainties            & 73 & 49 \\ 
$\chi^2$-contribution from logarithmic correction            & -129 & 48 \\ \hline \hline
Data set                                 & \multicolumn{2}{|c|}{partial $\chi^2 / n_{\rm dof}$}           \\ \hline
NC DIS HERA I  combined $e^-p$          & 108  / 145      & 108  / 145  \\
NC DIS  HERA I combined $e^+p$          & 419  / 379      & 419  / 379  \\
CC DIS HERA I combined $e^-p$           &  26  /  34      &  26  /  34   \\
CC DIS HERA I combined $e^+p$           &  39  /  34      &  41  /  34   \\
$c\bar{c}$ DIS HERA combined            &  78  /  52      &  47  /  52   \\
$b\bar{b}$ DIS ZEUS Vertex                  &  16  /  17      &  12  /  17   \\
LHCb $D^0$     &  68  /  38      &  17  /  30   \\
LHCb $D^{+}$   &  53  /  37      &  18  /  29   \\
LHCb $D^{*+}$     &  50  /  31      &  19  /  22   \\
LHCb $D_s^{+}$     &  24  /  28      &  11  /  20   \\
LHCb $\Lambda_c^{+}$  &   5.3 /  6       &   4.9 /   3\\
LHCb $B^{+}$   &  99  / 135      &  81  / 108\\
LHCb $B^0$     &  66  /  95      &  35  / 76\\
LHCb $B_s^0$     &  78  /  75      &  23  / 60 \\ \hline
\end{tabular}
\end{center}
\caption{
  \label{tab:chi}
  The global and partial $\chi^2$ values for the data sets used in the analysis of HERA and LHCb measurements.  
}
\end{table}

\begin{figure}[htbp]
\center
\epsfig{file=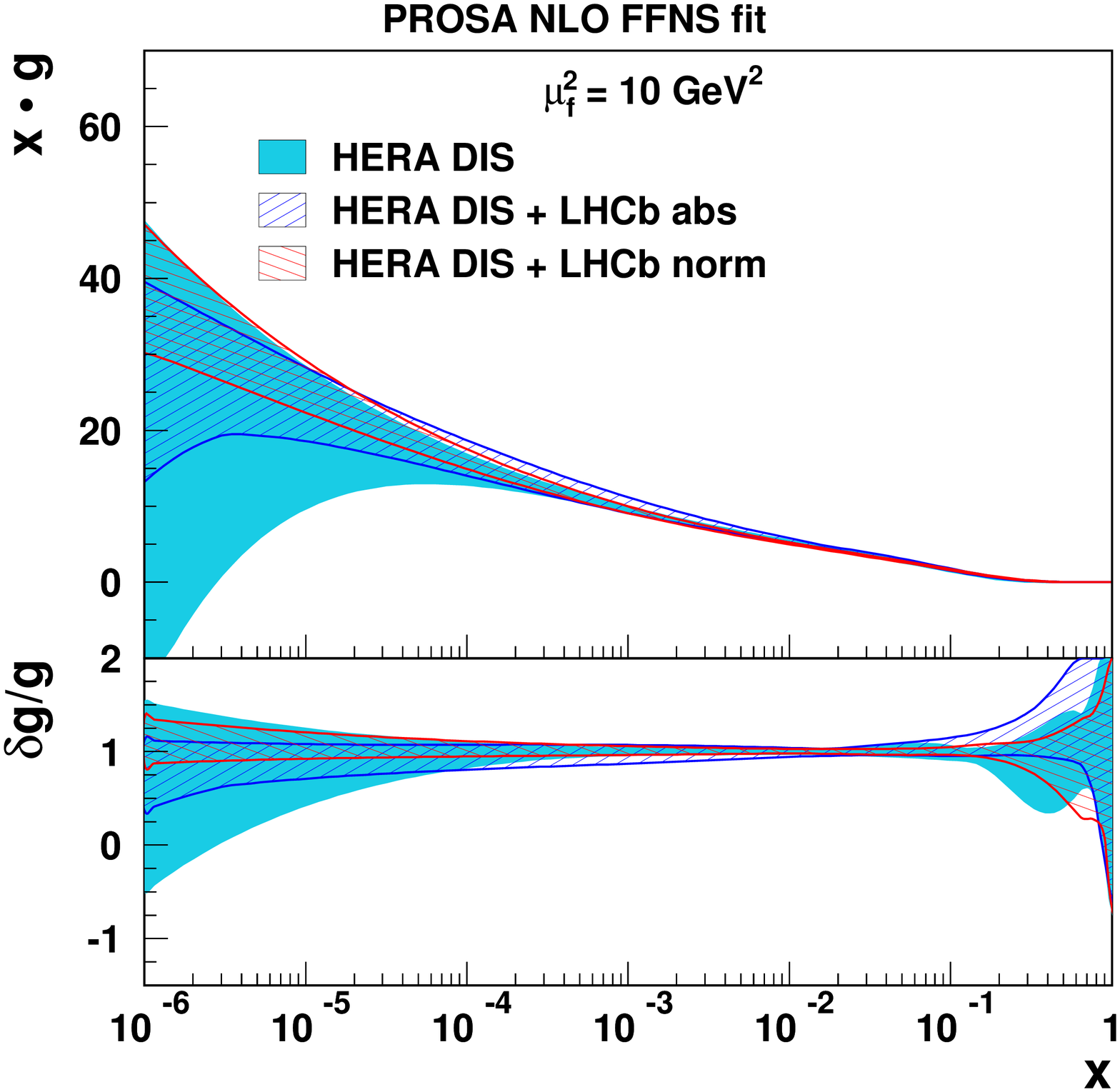,width=0.48\textwidth}
\epsfig{file=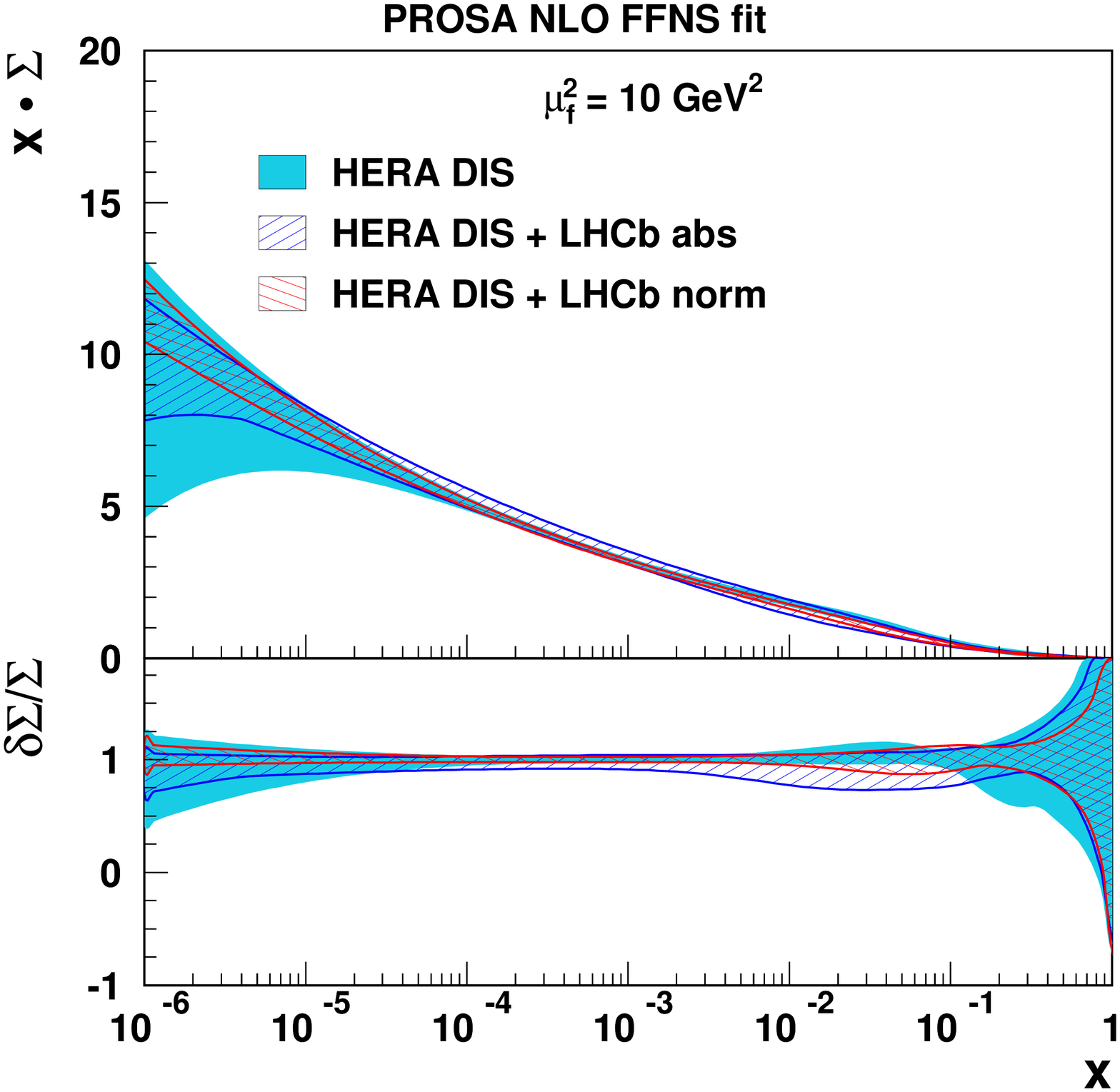,width=0.48\textwidth}\\
\epsfig{file=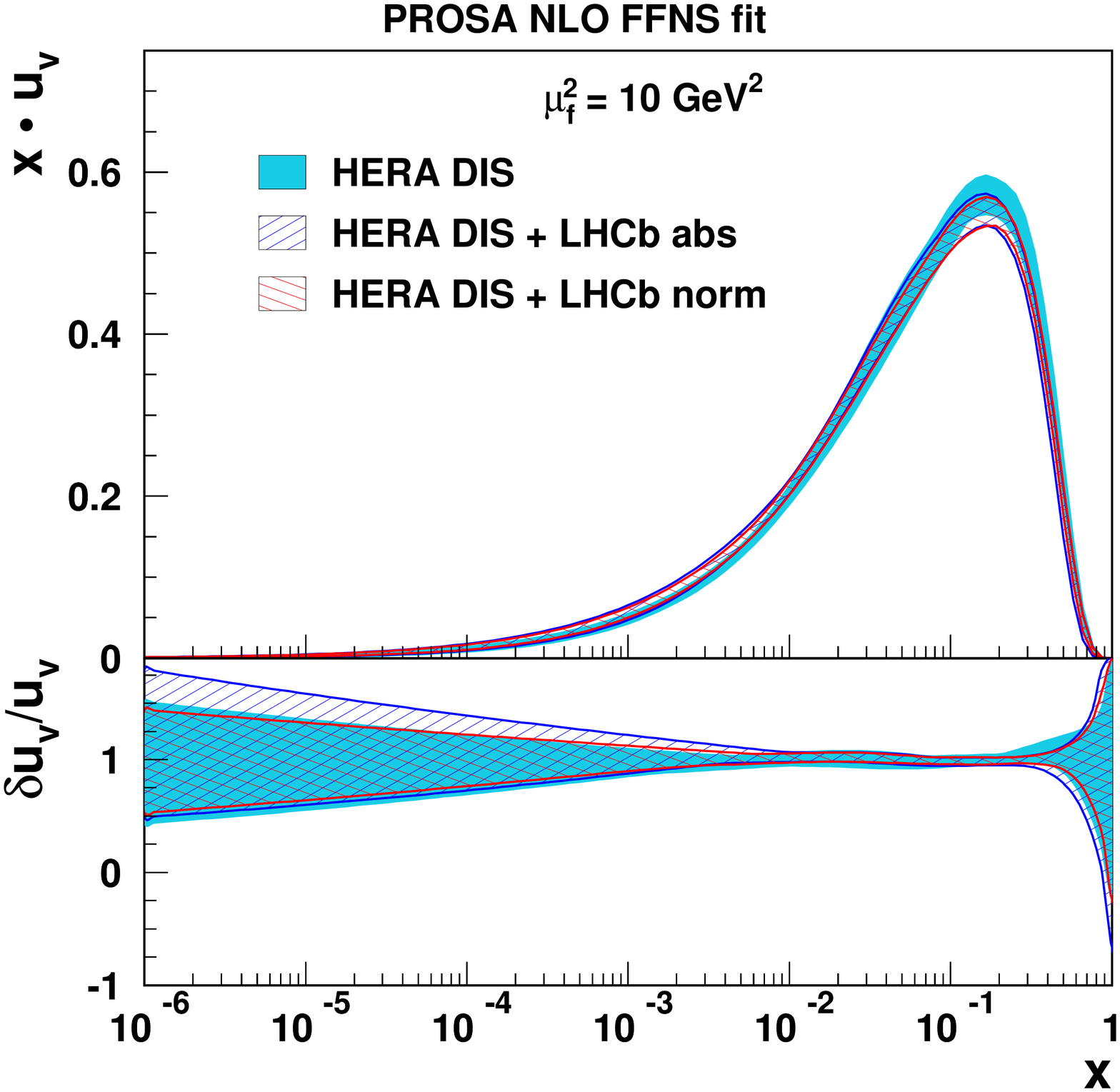,width=0.48\textwidth}
\epsfig{file=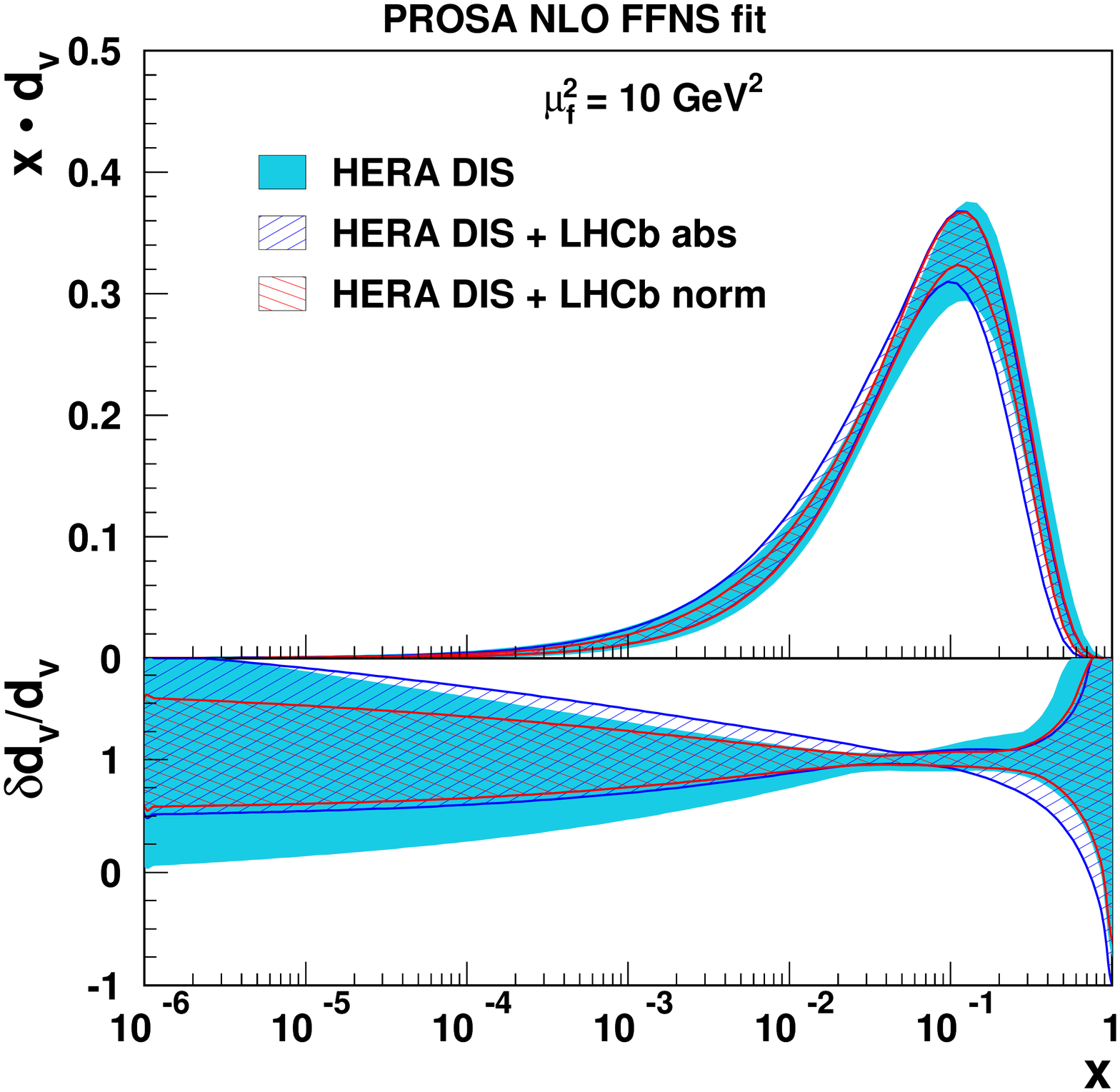,width=0.48\textwidth}
\caption{The gluon (top left), the sea-quark (top right), the $u$-valence quark (bottom left) and the $d$-valence quark (bottom right) 
distributions represented at $\mu_f^2 = 10$ GeV$^2$, as obtained in the QCD analysis of the HERA only data (light shaded band) and 
HERA and LHCb measurements and their relevant uncertainties. The sea-quark distribution is defined as $\Sigma=2 \cdot (\bar{u}+\bar{d}+\bar{s})$. 
The results of the fit using absolute or normalised LHCb 
measurements are shown by different hatches. The widths of the bands represent the total uncertainties.}
\label{fig:pdf_abs_norm} 
\end{figure}

The resulting gluon, valence-quark and sea-quark distributions with their total uncertainties are 
presented at $\mu_f^2=10$ GeV$^2$ in Fig.~\ref{fig:pdf_abs_norm} and compared to the result of the fit, based on 
solely HERA measurements of inclusive and heavy-flavour DIS. 
The uncertainties on the gluon and sea-quark distributions at low $x$ 
are significantly reduced in both cases, using LHCb absolute or normalised heavy-quark production cross sections. 
In case of the variant of the fit based on normalised LHCb cross sections, the uncertainties are 
reduced by more than a factor of three at $x \sim 5\times 10^{-6}$, which is the edge of the 
sensitivity of the included measurements (Fig. \ref{kinematics}). Consistent results are obtained 
in the fit using the absolute cross sections, which is considered an important cross 
check of the self-consistency of the NLO theory description. 

%The fact that the absolute variant yields a result which is consistent with 
%the the normalised one, despite the conceptual difficulties with the scale 
%treatment, is considered to be a valuable cross-check of the self-consistency 
%of the NLO theory description. 

\begin{figure}[htbp]
\center
\epsfig{file=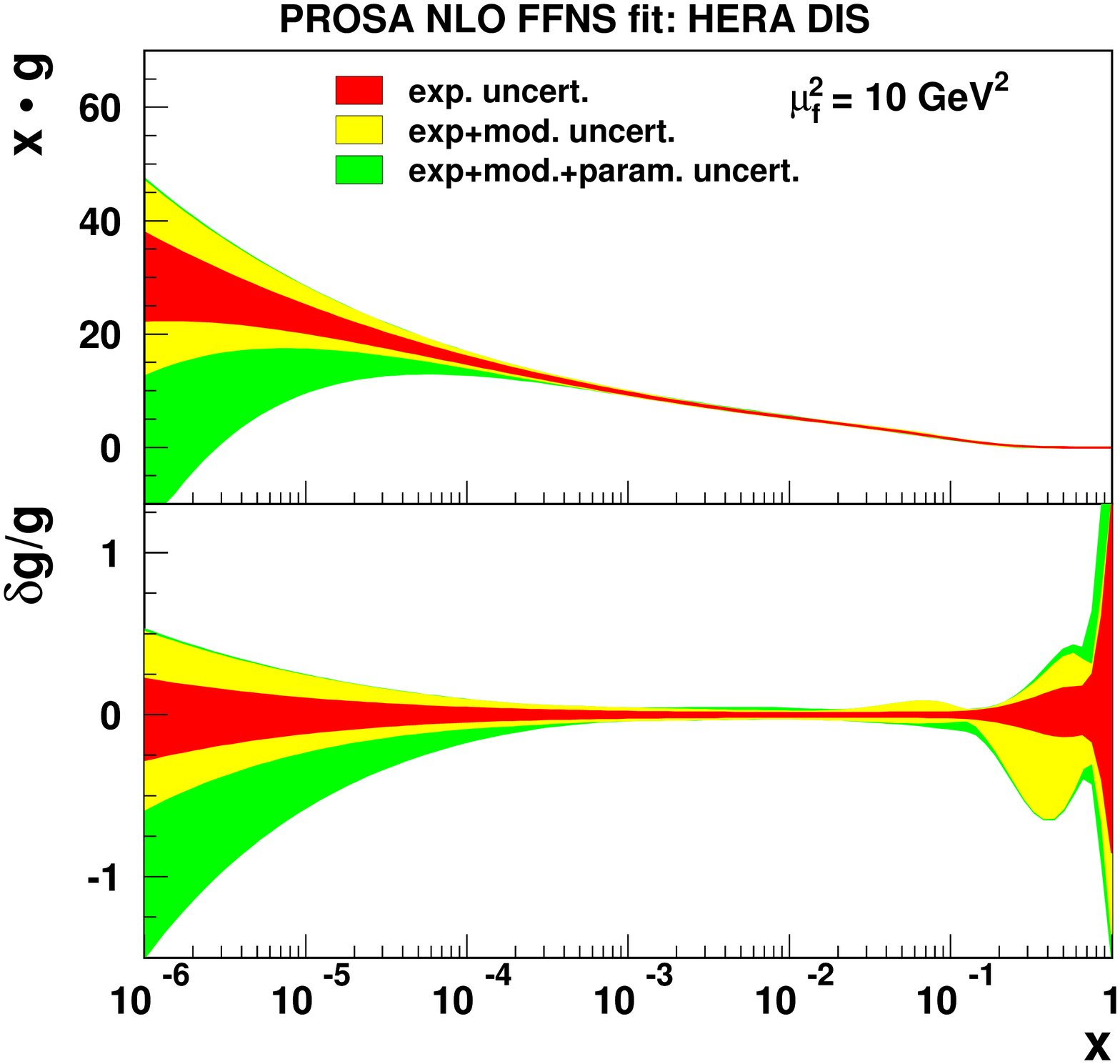,width=0.48\textwidth}\\
\epsfig{file=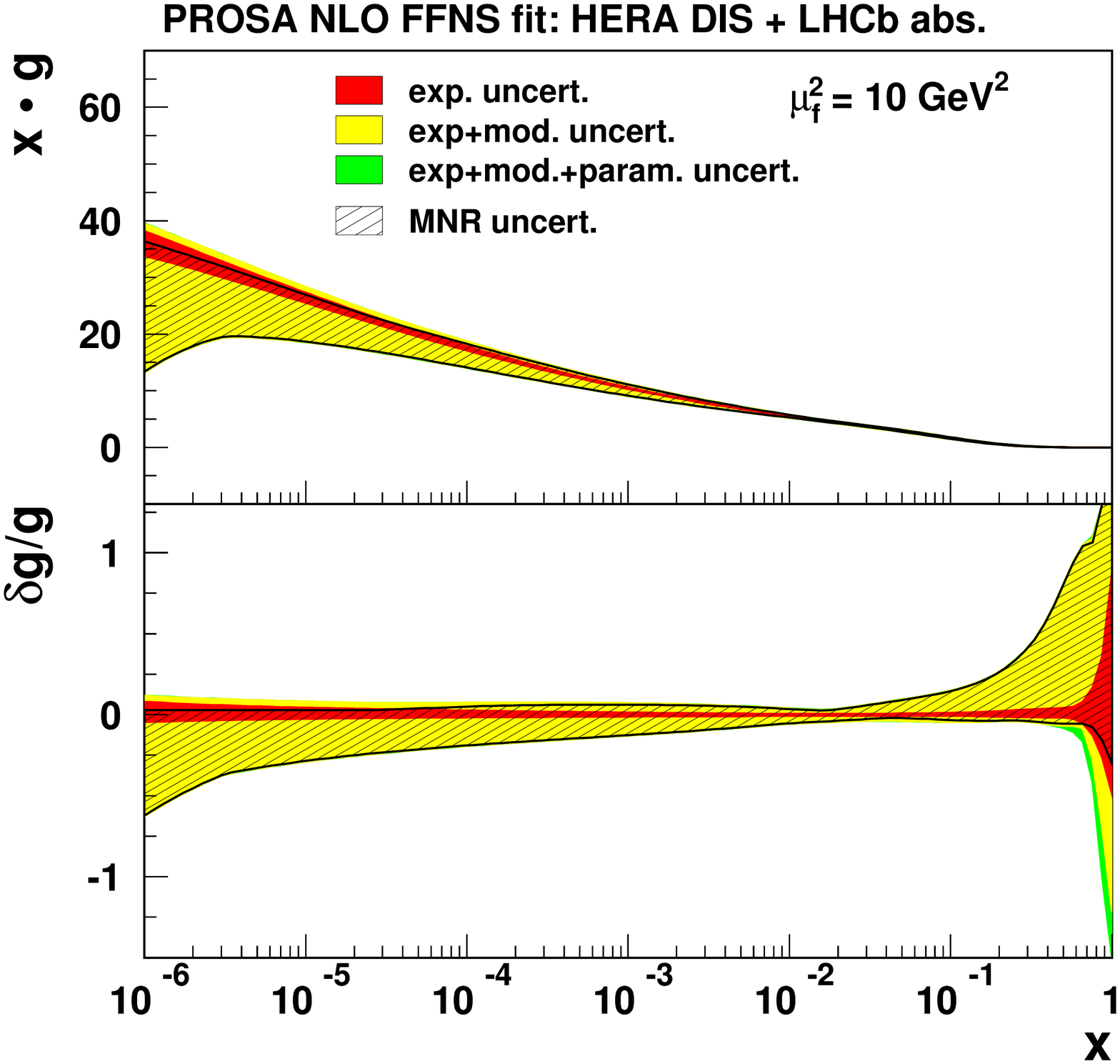,width=0.48\textwidth}
\epsfig{file=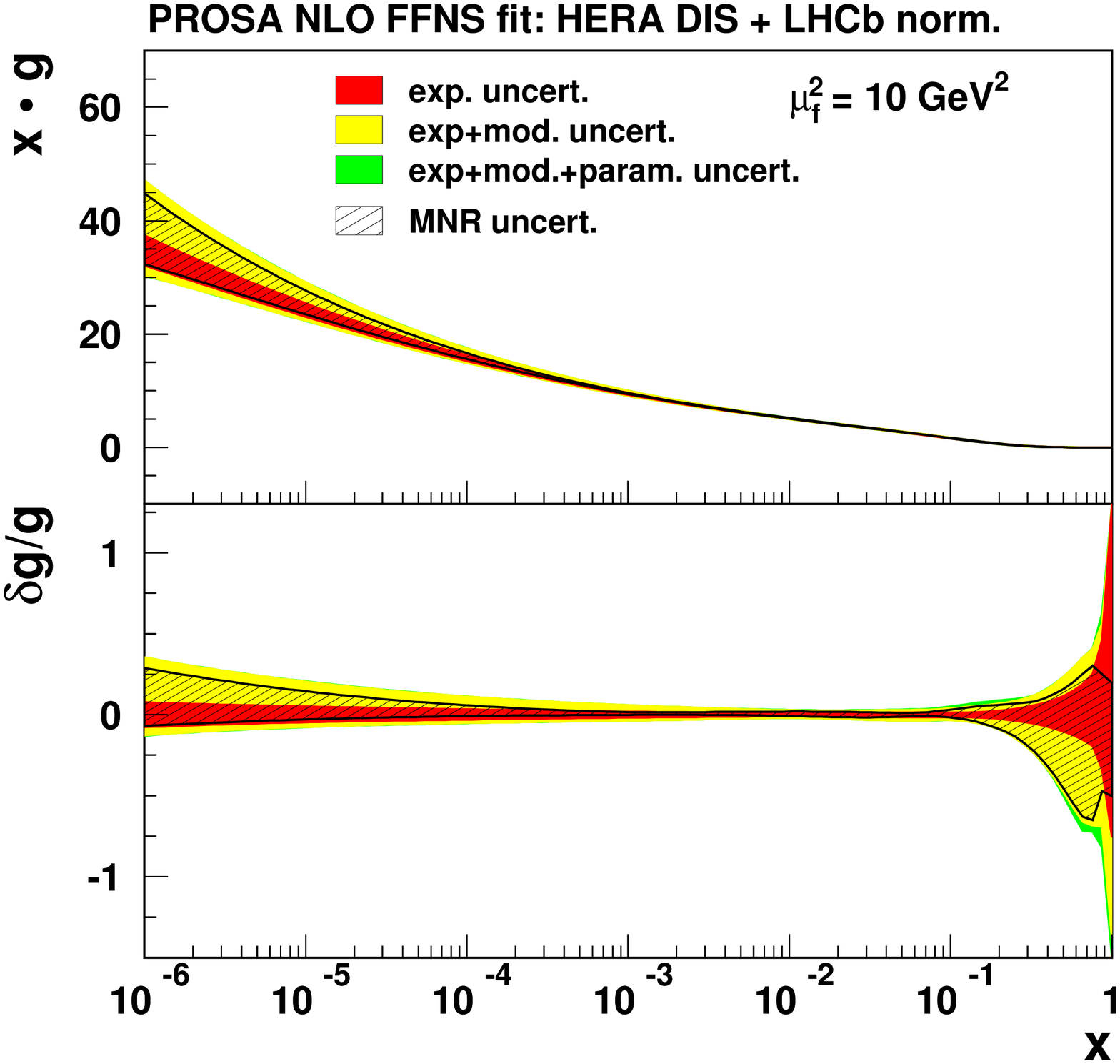,width=0.48\textwidth}
\caption{Individual contributions to the PDF uncertainties on the gluon distributions, obtained in QCD analyses
using HERA-only (upper panel), HERA+LHCb absolute (lower panel left) and HERA+LHCb normalised (lower panel right) cross sections of heavy-flavour production.}
\label{fig:pdf_error_contr} 
\end{figure}

The individual contributions of the experimental, model and parametrisation uncertainties for both cases 
of using the LHCb measurements are shown in Fig.~\ref{fig:pdf_error_contr} and compared to the result of 
the fit using only HERA data. The gluon distribution at low $x$ is constrained by the HERA 
measurements mostly via the sum rules and this results in large parametrisation uncertainties. 
%In this case the dominant source of the uncertainty is the variant of the parametrisation with free $D_{\overline U}$. 
Once the LHCb measurements are included in the QCD analysis, the gluon distribution is directly probed and 
the parametrisation dependence of the PDF is significantly reduced.

The main differences in the PDF uncertainties between the fits using the absolute and normalised LHCb measurements are 
caused by the MNR uncertainties. The variation of the pQCD scales in the prediction of the 
absolute cross section of heavy-flavour production in $pp$ collisions leads to significant changes in the 
normalisation of the cross section and represents the dominant uncertainty on the PDFs. The variations of the assumption on 
fragmentation parameters \cite{thesis} result in a smaller uncertainty, as compared to that due to the scale variations. 

In the case of the PDF fit using the normalised LHCb cross sections, the MNR uncertainty is strongly reduced, since variations 
of pQCD scales and of the fragmentation parameters do not significantly affect the shape of the $y$ distributions for 
heavy-flavour production. Therefore this is considered to be the primary result of this paper, while the 
consistency between the absolute and normalised variants is considered to be an important cross check.

%-----------------------------------------------------------------------------------------------------
\section{Conclusions}

The sensitivity of heavy-flavour production in $pp$ collisions to the low-$x$ gluon distribution was studied
in a comprehensive QCD analysis at NLO. The measurements of $c$- and $b$-hadron production cross sections at the 
LHCb experiment are included into a PDF fit together with inclusive and 
heavy-flavour production measurements 
in DIS at HERA. Since the bulk of the heavy-flavour data is close to the 
kinematic threshold, the fixed-flavour number scheme at next-to-leading order 
is used for the 
predictions of heavy-flavour production in $ep$ and $pp$ collisions. 
A significant reduction of the parametrisation uncertainty of the gluon 
distribution at very low $x$ is observed, as compared to the result of 
the PDF fit using only HERA DIS data. 

Two ways of using the LHCb measurements in the fit are studied. 
Although the absolute differential cross-section measurements contain more information, the resulting PDFs suffer 
from large theoretical uncertainty due to uncalculated higher-order corrections, estimated by the variation of 
the pQCD scales. By using only the rapidity shape information in the normalised cross sections for the final
result, this uncertainty is significantly reduced for the PDF extraction. 

The present analysis has illustrated the high potential of the LHCb measurements to constrain the gluon distribution 
at low $x$, and global PDF fits clearly can profit from the inclusion of such data. Precise measurements of normalised 
cross sections of heavy-flavour production in the forward kinematic range of the LHC therefore have a great potential 
to further improve the constraints on the PDFs.

In order to fully exploit the additional constraints from absolute 
LHC charm and beauty cross sections, a significant reduction of the theoretical uncertainties,
e.g. through threshold resummation and/or (partial) NNLO calculations with 
codes suitable for a usage in QCD analyses, is desirable.

\subsection*{Acknowledgements}

This work has been supported in part by the Helmholtz Association under contract numbers HGF-SO-072 and 
VH-HA-101. Supported in part by the EU-TMR Network Higgstools.

\clearpage
\appendix
%\section{Appendix}
\section{Appendix}
%\counterwithin{figure}{section}
\label{sec:Appendix}

\begin{figure}[tbp]
\epsfig{file=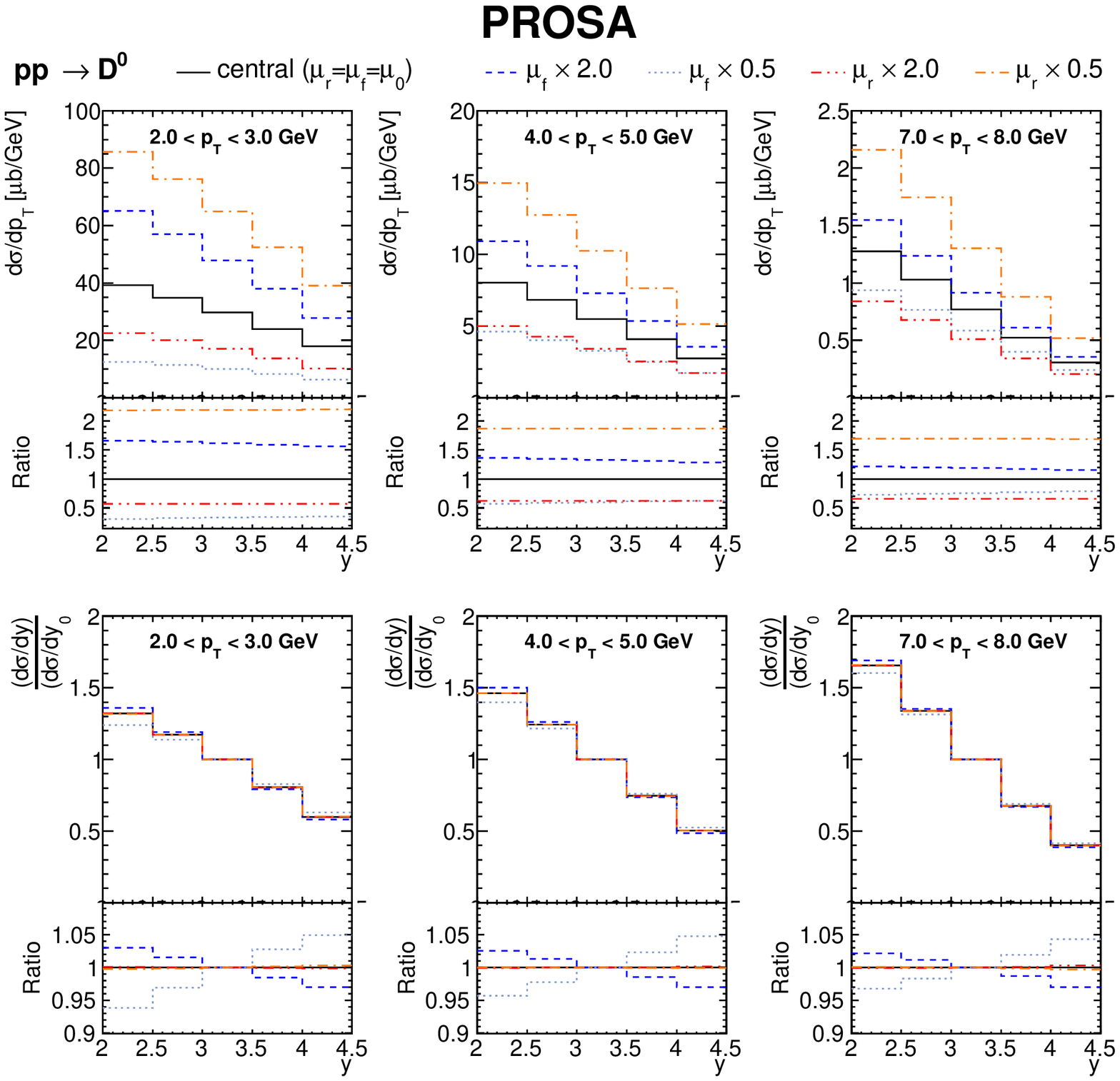,width=1.00\textwidth}
\caption{NLO QCD predictions for charm LHCb data with different scale choices for absolute (top) and normalised (bottom) cross sections. 
Lower inlets indicate the ratio of predictions to the central scale choice. The predictions are obtained by using the FFNS variant of MSTW 2008 PDFs\cite{Martin:2010db} with $N_f=3$; the charm mass is set to $m_c=1.5$ GeV.}
\label{fig:nlovar_c} 
\end{figure}

\begin{figure}[tbp]
\epsfig{file=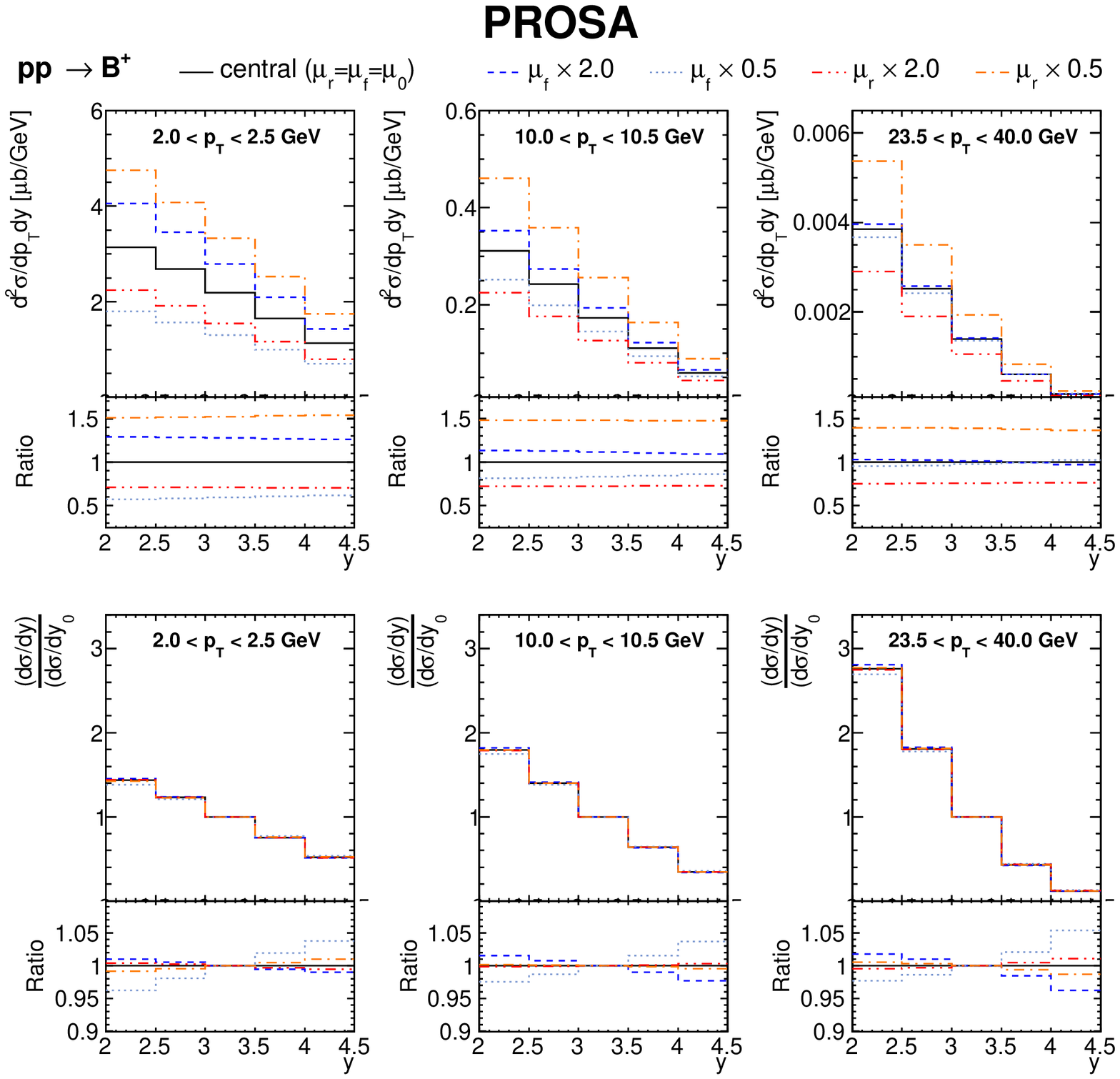,width=1.00\textwidth}
\caption{NLO QCD predictions for beauty LHCb data with different scale choices for absolute (top) and normalised (bottom) cross sections. 
Lower inlets indicate the ratio of predictions to the central scale choice. The predictions are obtained using the FFNS variant of MSTW 2008 PDFs\cite{Martin:2010db} with $N_f=3$; the beauty mass is set to $m_b=4.5$ GeV.}
\label{fig:nlovar_b} 
\end{figure}

\begin{table}[h]
\begin{center}
\renewcommand*{\arraystretch}{1.25}
\begin{tabular}{|l|c|c|} \hline
 Parameter   & absolute  &  normalised  \\ 
  \hline
  $B_g$ & $-0.14 \pm 0.07$& $-0.08 \pm 0.10$  \\ 
  $C_g$ & $6.83 \pm 0.31$& $5.23 \pm 0.34$  \\ 
  $A'_{g}$ & $1.74 \pm 0.22$& $1.29 \pm 0.32$  \\ 
  $B'_{g}$ & $-0.19 \pm 0.04$& $-0.16 \pm 0.05$  \\ 
%  $C'_{g}$ & $ 25.00 $& $ 25.00 $  \\ 
  $B_{u_{\textrm{v}}}$ & $0.668 \pm 0.020$& $0.649 \pm 0.021$  \\ 
  $C_{u_{\textrm{v}}}$ & $4.99 \pm 0.23$& $4.98 \pm 0.23$  \\ 
  $E_{u_{\textrm{v}}}$ & $12.2 \pm 2.4$& $13.5 \pm 2.7$  \\ 
  $B_{d_{\textrm{v}}}$ & $0.93 \pm 0.09$& $0.96 \pm 0.09$  \\ 
  $C_{d_{\textrm{v}}}$ & $5.50 \pm 0.56$& $5.59 \pm 0.55$  \\ 
  $C_{\overline {\textrm{U}}}$ & $1.63 \pm 0.21$& $1.63 \pm 0.24$  \\ 
  $A_{\overline {\textrm{D}}}$ & $0.173 \pm 0.007$& $0.158 \pm 0.007$  \\ 
  $B_{\overline {\textrm{D}}}$ & $-0.146 \pm 0.006$& $-0.155 \pm 0.007$  \\ 
  $C_{\overline {\textrm{D}}}$ & $10.4 \pm 2.5$& $15.1 \pm 4.2$  \\ 
%  $\alpha_s$ & $ 0.1059 $& $ 0.1059 $  \\ 
%  $f_s$ & $ 0.3100 $& $ 0.3100 $  \\ 
  $m_c$ & $1.709 \pm 0.024$& $1.257 \pm 0.014$  \\ 
  $m_b$ & $4.67 \pm 0.08$& $4.19 \pm 0.13$  \\ 
%  fp1c & $\textcolor{blue}{ 4.400 }$& $\textcolor{blue}{ 4.400 }$  \\ 
%  fp1b & $\textcolor{blue}{ 11.00 }$& $\textcolor{blue}{ 11.00 }$  \\ 
%  $A_{f}^{c}$ & $0.434 \pm 0.027$& $ 1.0 $  \\ 
%  $A_{f}^{b}$ & $0.0688 \pm 0.0035$&$ 1.0 $  \\ 
%  $A_{r}^{c}$ & $0.197 \pm 0.019$& $ 1.0 $  \\ 
%  $A_{r}^{b}$ & $0.112 \pm 0.016$& $ 1.0 $  \\ 
% square root
  $A_{f}^{c}$ & $0.659 \pm 0.020$& $ 1.0 $  \\ 
  $A_{f}^{b}$ & $0.262 \pm 0.007$&$ 1.0 $  \\ 
  $A_{r}^{c}$ & $0.444 \pm 0.021$& $ 1.0 $  \\ 
  $A_{r}^{b}$ & $0.335 \pm 0.024$& $ 1.0 $  \\ 
  \hline
    \end{tabular}
  \end{center}
\caption{
  \label{tab:fitpar}
  The fitted parameters for the NLO QCD analysis using HERA and LHCb measurements. The value of strong 
coupling $\alpha_S(m_Z)^{N_f=3} =0.1059$ is used. The listed uncertainties correspond to those 
associated to the experimental measurements used in the fit. Uncertainties are not quoted for parameters 
that are fixed.
}
\end{table}

\end{document}